\documentclass[12pt]{iopart}
\usepackage{amsfonts}
\usepackage{iopams}
\usepackage{amsthm}
\usepackage{braket}
\usepackage{graphicx}
\usepackage{hyperref}
\usepackage{color}
\usepackage{amssymb}
\usepackage{dsfont}

\begin{document}

\title{Numerical identification and gapped boundaries of Abelian fermionic topological order}

\author{Nick Bultinck}

\address{Department of Physics, University of California, Berkeley, CA 94720, USA}

\vspace{10pt}
\begin{indented}
\item[]November 2019
\end{indented}

\begin{abstract}
In this work we consider general fermion systems in two spatial dimensions, both with and without charge conservation symmetry, which realize a non-trivial fermionic topological order with only Abelian anyons. We address the question of precisely how these quantum phases differ from their bosonic counterparts, both in terms of their edge physics and in the way one would identify them in numerics. As in previous works, we answer these questions by studying the theory obtained after gauging the global fermion parity symmetry, which turns out to have a special and simple structure. Using this structure, a minimal scheme is outlined for how to numerically identify a general Abelian fermionic topological order, without making use of fermion number conservation. Along the way, some subtleties of the momentum polarization technique are discussed. Regarding the edge physics, it is shown that the gauged theory can have a (bosonic) gapped boundary to the vacuum if and only if the ungauged fermion theory has a gapped boundary as well.
\end{abstract}

%
%
%
%
%

\section{Introduction}

Since the experimental discovery of the fractional quantum Hall effect, a tremendous effort has been devoted to the study of topologically ordered phases that can be realized in strongly correlated quantum many-body systems. In the past decade, substantial progress has been made in the characterization of topologically ordered phases in spin or boson systems, and a unified algebraic framework called `modular tensor categories' has been identified to describe these phases \cite{Kitaev}. From a physical point of view, a modular tensor category simply describes how the anyonic excitations fuse and braid with each other. Phrasing the properties of topologically ordered phases in the rigid algebraic language of modular tensor categories has allowed theorists to make significant progress in the study of bosonic topological phases.

Not only has our theoretical understanding of topologically ordered phases vastly improved, in recent years many important results have been obtained on how to numerically identify the type of topological order realized by a particular microscopic Hamiltonian. For example, it was realized that a non-trivial topological order leaves an imprint on the entanglement entropy of a spatial region in the ground state via the `topological entanglement entropy' term \cite{KitaevPreskill,LevinWen}. A later refinement of the topological entanglement entropy showed that the complete spectrum of the reduced density matrix of a spatial region contains information about the universal edge physics of the topological phase \cite{LiHaldane}. When the system of interest is put on a torus, it will necessarily have a ground state degeneracy if a a non-trivial topological order is realized \cite{WenNiu}. In Refs. \cite{ZhangGrover,Cincio}, a useful basis for the ground state subspace, the so-called Minimally-Entangled State (MES) basis, was identified and it was shown that this basis can be used to obtain the $S$-matrix, which contains information about the anyon braiding statistics, and the $T$-matrix, which gives access to the chiral central charge and the topological spins of the anyons. Later works showed that these MES also give access to the topological spins of the anyons via a quantity called the `momentum polarization' \cite{ZaletelMong,TuZhang}.

In this work, we focus on fermion systems with Abelian topological order, meaning that the anyons form an Abelian group under fusion. In the algebraic languague, the most important difference between bosonic and fermionic topological orders is that the latter don't satisfy the same strict requirement of modularity as bosonic systems do. We explain this in more detail in the main text, where the algebraic frameworks for both bosonic and fermionic Abelian topological orders are reviewed. A fermionic system necessarily has a fermion parity symmmetry, and a useful tool in the study of fermionic topological orders is to gauge this symmetry \cite{Kitaev} (the approach of gauging a global symmetry has also proven to be very useful in the study of symmetry-protected phases \cite{LevinGu}). Importantly, since the microscopic fermion becomes a non-local gauge charge after gauging, the resulting theory is purely bosonic. Below, two main questions regarding Abelian fermionic topological orders (AfTO) are addressed: (1) How does one uniquely identify the most general AfTO in numerics?, and (2) Does the existence of a gapped edge for the ungauged fermionic topological order imply the existence of a gapped edge for the gauged topological order and vice versa?

Regarding the first question, we note that a lot of the above mentioned numerical techniques for identifying a topological order have been successfully applied to fractional quantum Hall systems and fractional Chern insulators \cite{Neupert,ShengGu,Regnault,ZaletelMong,Rezayi,Grushin}, which are of course fermionic in nature. These approaches relied crucially on the presence of a fermion number conservation symmetry, which allows for the definition of a Hall conductivity $\sigma_{xy}$. In fractional quantum Hall systems one knows the value of $\sigma_{xy}$ exactly from Galilean invariance (if there is no disorder), and in lattice systems it can be computed numerically as the many-body Chern number \cite{Niu,ShengHaldane}, or from the entanglement eigenvalues by an adiabatic flux insertion procedure \cite{Zaletel,Grushin} (see also Ref. \cite{Alexandradinata}). In this work, we outline an alternative numerical detection scheme which does not rely on computing the Hall conductance, and which can identify the most general Abelian fermionic topological order. In formulating this numerical scheme, we will make heavy use of the special structure of gauged AfTOs. Another important ingredient for our detection scheme is the momentum polarization technique \cite{ZaletelMong,TuZhang}, and we point out some properties of this numerical probe which --to the best of our knowledge-- have not been discussed previously in the literature and which are relevant for fermionic systems.

As for the second question regarding the edge physics of topological phases, we will show that the bosonic theory which is obtained by gauging fermion parity in an AfTO can have a (conventional, bosonic) gapped edge with the trivial vacuum if and only if the ungauged fermionic theory can have a gapped edge as well. To show that gauging fermion parity does not change whether a system admits a gapped edge or not, we use the bulk-boundary correspondence formulated in terms of Lagrangian subgroups as introduced in Ref. \cite{Levin}.

For completeness, we also mention some previous works which have studied fermionic topological orders and are relevant for the present work. Refs. \cite{GuWangWen,TianLan1,TianLan2} worked out an algebraic framework for general fermionic topological orders (both Abelian and non-Abelian), which is a generalization of the modular tensor categories for bosonic topological orders. Some ideas of these works will be used below. Refs. \cite{Belov,Cano} have studied Abelian fermionic topological orders with fermion number conservation symmetry using multi-component U$(1)$ Chern-Simons theories. The algebraic approach adopted in this work agrees with the U$(1)$ Chern-Simons approach of Refs. \cite{Belov,Cano} where the results overlap. And finally, Ref. \cite{Wang} has studied general gauged fermionic topological orders from a mathematical perspective, and conjectured that Kitaev's $16$-fold way \cite{Kitaev} has a natural generalization to the most general fermionic topological order. This being said, let us now turn to a short review of the algebraic framework behind bosonic and fermionic Abelian topological order.

\section{Abelian bosonic topological order}

In this section, we briefly review the properties of Abelian bosonic topological orders (AbTO) that are relevant for the main discussion below. For more details, the reader is referred to Ref. \cite{Kitaev}. In mathematical terms, an AbTO is equivalent to an Abelian modular tensor category. To specify such an Abelian modular tensor category, we need to provide a list $\mathcal{A}=\{a,b,c,\dots\}$ of $N$ anyon types, together with the corresponding Abelian fusion rules $a\times b = \sum_c N_{a,b}^c c$ such that $N_{a,b}^c\in \{0,1\}$, and topological spins $\theta_a$ \footnote{This is a slight abuse of terminology which is common in the literature. If we write $\theta_a=e^{2\pi i h_a}$, then it is more appropriate to call $h_a$ the topological spin. However, in this work it will be more convenient to simply refer to $\theta_a$ as the topological spin.}. Every AbTO has a unique trivial anyon, denoted as $1$, which has the properties $1\times a = a$ and $\theta_1=1$.  Often, we will write the fusion of two anyons $a$ and $b$ simply as $ab = a\times b$. In principle, we also need to provide the $F$-symbols, but they will not be important here so we omit them. We denote the braiding phase associated with moving an anyon $b$ counter-clockwise around anyon $a$ as $M_{a,b}$. By definition, the braiding phases are symmetric: $M_{a,b}=M_{b,a}$. The ribbon identity allows $M_{a,b}$ to be expressed in terms of the topological spins:

\begin{equation}
M_{a,b}=\frac{\theta_{ab}}{\theta_a\theta_b}
\end{equation}
The $S$ matrix is related to the braiding phases by $S_{a,b}=M_{a,b}\mathcal{D}^{-1}$, where the total quantum dimension $\mathcal{D}$ of an Abelian modular tensor category is given by $\mathcal{D}=\sqrt{N}$. Modularity requires that $S^\dagger S=\mathds{1}$. 

An important connection between the bulk anyons and the boundary theory of an AbTO is given by the following relation \cite{Kitaev}:

\begin{equation}\label{kit}
\frac{1}{\sqrt{N}}\sum_a \theta_a = e^{2\pi i c_-/8}\, ,
\end{equation}
where $c_-$ is the chiral central charge of the boundary theory. This implies that the bulk anyons determine the boundary chiral central charge up to a multiple of $8$. This is the best one can do, as there exists an invertible bosonic topological phase, the so-called $E_8$ state \cite{KitaevE8}, which has no anyons but a chiral edge with $c_-=8$. One can thus always stack an $E_8$ state on top of the system of interest, which does not affect the bulk anyons but changes the boundary chiral central charge by $8$.

\subsection{Gapped boundaries}\label{GappedB}

As was shown by Levin, an AbTO admits a gapped edge iff (1) $c_-=0$ and (2) the bulk anyons have a bosonic Lagrangian subgroup \cite{Levin}. A bosonic Lagrangian subgroup $\mathcal{L}_b$ is a subgroup of anyons which have the following properties (see also Refs. \cite{JuvenWang1,JuvenWang2}):

\begin{itemize}
\item[1)] Every anyon in $\mathcal{L}_b$ has trivial topological spin,
\item[2)] All anyons in $\mathcal{L}_b$ braid trivially with each other,
\item[3)] Every anyon which is not in $\mathcal{L}_b$ braids non-trivially with at least one anyon in the Lagrangian subgroup.
\end{itemize}
Using Eq.~(\ref{kit}), we will show that the existence of a Lagrangian subgroup implies that the chiral central charge is a multiple of $8$. It thus follows that an AbTO allows for a gapped edge iff it has a Lagrangian subgroup, and the separate requirement of zero chiral central charge is redundant, provided that we are allowed to stack $E_8$ states on top of our system. Combined with the results of Ref. \cite{Levin2}, this implies that every AbTO with a Lagrangian subgroup has a string-net representation \cite{stringnet}.

To derive $c_-=0$ mod $8$ from the existence of a bosonic Lagrangian subgroup $\mathcal{L}_b$, we write the AbTO as $\mathcal{A}=\{\mathcal{L}_b,\mathcal{L}_b\times n_1, \mathcal{L}_b\times n_2,\dots\}$, where $n_i$ are a set of arbitrary anyons not in $\mathcal{L}_b$. Starting from Eq.~(\ref{kit}), we can now do the following manipulations:

\begin{eqnarray}
e^{2\pi ic_-/8} & = & \frac{1}{\sqrt{N}}\sum_{a\in\mathcal{A}} \theta_a \\
 & = &  \frac{1}{\sqrt{N}}\sum_{l \in \mathcal{L}_b}\theta_l +  \frac{1}{\sqrt{N}} \sum_{n_i}\sum_{l \in \mathcal{L}_b}\theta_{n_il}  \\
 & = &  \frac{1}{\sqrt{N}}\sum_{l \in \mathcal{L}_b}\theta_l +  \frac{1}{\sqrt{N}} \sum_{n_i}\theta_{n_i} \sum_{l \in \mathcal{L}_b}\theta_{l}M_{l,n_i} \\
 & = &  \frac{N_{L}}{\sqrt{N}}+  \frac{1}{\sqrt{N}} \sum_{n_i}\theta_{n_i} \sum_{l \in \mathcal{L}_b}M_{l,n_i} \label{sum}\\
 & = & \frac{N_{L}}{\sqrt{N}}\, ,\label{Lb}
\end{eqnarray}
where $N_{L}$ is the number of anyons in $\mathcal{L}_b$. In the third line we have used the ribbon identity, and in the fourth line we relied on the definition of a bosonic Lagrangian subgroup wich states that $\theta_l=1$ if $l\in \mathcal{L}_b$. To see why the second term in~(\ref{sum}) is zero, note that $M_{l_1,n_i}M_{l_2,n_i}=M_{l_1l_2,n_i}$, such that $M_{l,n_i}$ for fixed $n_i$ forms a representation of $\mathcal{L}_b$. Because there is at least one $l\in\mathcal{L}_b$ for which $M_{l,n_i}\neq 1$, this representation cannot be the trivial representation. Schur's orthogonality relations then imply that the sum of $M_{l,n_i}$ over all $l\in\mathcal{L}_b$  is zero. We have thus obtained the desired result that $c_-=0$ mod $8$ if there exists a Lagrangian subgroup. As a side-result, we also found that $N_{L}^2=N$, such that only AbTOs where the number of anyons is a square number can have a Lagrangian subgroup (this relation between $N_{L}$ and $N$ was also obtained previously in Ref. \cite{Levin}).

\section{Abelian fermionic topological order}

The main difference between a fermionic topological order (fTO) and a bTO, is that a fTO has a distinguished fermion excitation $f$ with the properties $f^2=f\times f=1$ and $\theta_f=-1$, which is `transparent', i.e. it braids trivially with all other particles. This is not allowed in a bTO because the existence of such a particle is a violation of modularity. In Refs. \cite{Cano,Cheng}, it was shown that every Abelian fermionic topological order (AfTO) $\mathcal{A}_f$ can be written as

\begin{equation}\label{prod}
\mathcal{A}_f = \mathcal{A}_b\times \{1,f\}\, ,
\end{equation}
where $\mathcal{A}_b$ is an AbTO. This result also follows from corollary A.19 of Ref. \cite{Drinfeld}. Note that the above factorization property does not hold for non-Abelian fTO, as is known from explicit counter examples \cite{TianLan1,TianLan2}. 

It is important to keep in mind that in general, the factorization in Eq.~(\ref{prod}) is not unique. In particular, let us define the homomorphism $\beta: \mathcal{A}_b\rightarrow \mathbb{Z}_2$, where $\beta(a)$ takes values in $\{0,1\}$. The fact that $\beta$ is a homomorphism then implies that $\beta(a b) = \beta(a)+\beta(b)$ mod $2$. Using $\beta$, we can rewrite the factorization in Eq. \ref{prod} in a different way as $\mathcal{A}_f = \mathcal{A}_b^\beta\times\{1,f\}$, where $\mathcal{A}_b^\beta = \{a f^{\beta(a)}|a\in\mathcal{A}_b\}$. So the number of different factorizations is given by the number of homomorphisms from $\mathcal{A}_b$ to $\mathbb{Z}_2$.

A fermionic system necessarily has fermion parity symmetry. This implies that we can introduce a corresponding fermion parity flux or $\pi$-flux into the system. In the absence of U$(1)$ fermion number symmetry, there is no unique way of defining such a fermion parity flux. In particular, for a given parity flux $\phi_i$ we can always obtain a different parity flux by attaching an anyon in $\mathcal{A}_f$ to it. This of course assumes that the parity flux cannot `absorb' the anyons in $\mathcal{A}_f$, i.e. that $\phi_i \times a\neq \phi_i$ for $a \in \mathcal{A}_f$. While it is possible for $\phi_i$ to absorb the transparent fermion $f$, we will show in the next section that the parity fluxes cannot absorb the anyons in $\mathcal{A}_b$. The case where the fermion parity fluxes can absorb $f$ occurs in superconducting systems where a Majorana mode binds to the parity flux, like in the weak pairing phase of spinless $p-$wave superconductors  \cite{Volovik,ReadGreen}. When this happens, the parity fluxes are non-Abelian defects \cite{Ivanov}.

An important property of fermion parity fluxes is that they have a well-defined topological spin. This is different from $\mathbb{Z}_2$ fluxes in bosonic systems, which have a topological spin that is only defined up to a minus sign. The reason is that bosonic systems admit trivial particles with both even and odd $\mathbb{Z}_2$ charge. So if we attach such a trivial charge one object to a $\mathbb{Z}_2$ flux, we don't change the superselection sector, but we do change its topological spin by $-1$. In fermionic systems, however, all charge one particles have topological spin $\theta_f=-1$, so by attaching them to a parity flux we don't change the topological spin.

\subsection{Gapped boundaries}\label{GappedF}

Similarly to the bosonic case, Levin showed that an AfTO admits a gapped edge if and only if $c_-=0$, and there exists a fermionic Lagrangian subgroup $\mathcal{L}_f$, which is defined to have the following properties \cite{Levin}:

\begin{itemize}
\item[1)] All anyons in $\mathcal{L}_f$ braid trivially with each other,
\item[2)] Every anyon in $\mathcal{A}_b$ which is not in $\mathcal{L}_f$ braids non-trivially with at least one anyon in the Lagrangian subgroup.
\end{itemize}
The only difference between the definitions of $\mathcal{L}_b$ and $\mathcal{L}_f$ is that in the latter we do not require the anyons in $\mathcal{L}_f$ to have trivial topological spin. Note, however, that because $\theta_l^2=M_{l,l}=1$ for all $l\in\mathcal{L}_f$, it follows that the definition of a fermionic Lagrangian subgroup only leaves a sign ambiguity in the topological spins of the anyons in $\mathcal{L}_f$.

\section{Gauging fermion parity: modular extensions of an AfTO}\label{sec:gauging}

There exists a well-defined microscopic prescription to gauge the fermion parity symmetry in any fermionic lattice Hamiltonian \cite{Kogut}. After gauging, the system realizes a bosonic topological order in the bulk. This is because the gauging procedure promotes the parity fluxes to deconfined anyonic excitations, which braid non-trivially with $f$. This implies that the gauged theory is modular, and can be realized in a bosonic system. In mathematical terms, a GfTO is called a `modular extension' of the original fTO \cite{TianLan1,TianLan2}. In this section, we will show that for AfTO such modular extensions have a special structure. For this we consider the process where one creates an $a-\bar{a}$ anyon pair from the vacuum and braids one of them, say $a$, around a fermion parity flux. It is well-known from bosonic symmetry-enriched topological orders that braiding around a symmetry defect $g$ can permute the anyon types \cite{Bombin,Barkeshli,Tarantino,Chen}. This means that after braiding an anyon $a$ around $g$, it is possible for $a$ not to come back as itself, but as a different anyon $\pi_g(a)$. After gauging, the parity fluxes become deconfined anyonic excitations. Because braiding of anyons cannot change the anyon type, this implies that after gauging the anyons $a$ and $\pi_g(a)$ have to be identified as the same anyon \cite{Barkeshli,Tarantino,Chen}. However, we now argue that this cannot happen for a fermion parity flux, i.e. for fermion parity we always have $\pi_\phi(a)=a$. The reason is that fermionic Hilbert spaces have a superselection rule which states that every physical state needs to have well-defined fermion parity \cite{Wick}. So if fermion parity could permute anyons, then starting from an excited state with some localized anyons it would be possible to create an orthogonal state by acting with fermion parity, but this clearly violates the superselection rule. 

As already anticipated above, we can now also argue that fermion parity fluxes cannot absorb anyons in $\mathcal{A}_b$. To see this, assume that a parity flux $\phi_i$ could absorb an anyon $a\in\mathcal{A}_b$. The only way this can happen consistently, is if all anyons $b$ with $M_{a,b}\neq 1$ get permuted when they braid with the parity flux. But as we just argued, this is impossible. Note that $f$ escapes this argument since it braids trivially with all anyons in $\mathcal{A}_f$.

When gauging a $\mathbb{Z}_2$ symmetry in bosonic systems, the trivial anyon $1$ before gauging splits into a trivial anyon and a non-trivial anyon after gauging. This is because the original symmetry-enriched topological order has trivial anyons with both even and odd $\mathbb{Z}_2$ charge. Under gauging, the trivial anyons with even charge remain trivial, but the trivial anyons with odd charge become the gauge charge anyons of the gauged theory. For fermionic systems, this does not happen because a fTO has no trivial anyons with odd fermion parity.

The above two arguments show that the anyons of a fTO do not get identified and do not split after gauging fermion parity, which implies that the original fTO is a subcategory of the GfTO. For AfTO, if $\mathcal{A}_f\times\{1,f\}$ is a subcategory of the GfTO, then $\mathcal{A}_b$ is obviously also a subcategory of the gauged theory. We can now use theorem 3.13 from Ref. \cite{Drinfeld}, which says the following: 
\\ \\
\textbf{Theorem (Ref. \cite{Drinfeld}).} Consider a MTC $\mathcal{K}$, and assume that it is a fusion subcategory of $\mathcal{C}$, i.e. $\mathcal{K}\subset\mathcal{C}$. If $\mathcal{C}$ is a MTC, then $\mathcal{C}=\mathcal{K}\times\mathcal{K}'$, where $\mathcal{K}'$ is also a MTC. 
\\ \\
Because $\mathcal{A}_b$ is modular, we can directly apply the above theorem to conclude that the GfTO takes the form

\begin{equation}\label{factor1}
GfTO = \mathcal{A}_b \times\{1,f,\phi,f\phi\}\, ,
\end{equation}
if the fermion parity fluxes are Abelian, and

\begin{equation}\label{factor2}
 GfTO = \mathcal{A}_{b}\times\{1,f,\phi\}\, ,
\end{equation}
if the fermion parity fluxes are non-Abelian. Both $\phi$ and $f\phi$ are anyons which correspond to deconfined fermion parity fluxes.

In the present context, it is not hard to prove the factorization of the GfTO, as given in Eqs.~(\ref{factor1}) and (\ref{factor2}), without invoking theorem 3.13 of Ref. \cite{Drinfeld}. To see this, note that since fermion parity fluxes $\phi_i$ cannot permute anyons, the process of making an $a-\bar{a}$ pair, braiding $a$ around $\phi_i$, and subsequently annihilating the anyon pair again is a well-defined adiabatic process for every parity flux $\phi_i$ and anyon $a\in\mathcal{A}_b$. Therefore, we can associate a Berry phase to it. Because there are no trivial particles with odd fermion parity, the Berry phases depend only on the anyon type and it makes sense to write them as $e^{i\gamma_i(a)}$ and $e^{i\gamma_i(af)}$ for every flux $\phi_i$ and $a\in\mathcal{A}_b$. The Berry phases satisfy the obvious properties $e^{i\gamma_i(a b)}=e^{\gamma_i(a)}e^{i\gamma(b)}$ and $e^{i\gamma_i(f)}=-1$. Because $\mathcal{A}_b$ is modular, we can use lemma 3.31 from Ref. \cite{Drinfeld}  (see also Ref. \cite{Barkeshli}, page 11), which states that for every function $e^{i\gamma_i(\cdot)}:\mathcal{A}_b\rightarrow U(1)$ that satisfies $e^{i\gamma_i(a b)}=e^{\gamma_i(a)}e^{i\gamma_i(b)}$, there exists a corresponding unique anyon $a_i\in \mathcal{A}_b$ such that

\begin{equation}\label{braidingthm}
e^{i\gamma_i(a)}=M_{a,a_i}\,,\;\;\forall a \in \mathcal{A}_b
\end{equation}
This implies that for every parity flux $\phi_i$, we can find an anyon $a_i\in\mathcal{A}_b$ such that $\phi_i \bar{a}_i$ braids trivially with all anyons in $\mathcal{A}_b$. Because the different $\phi_i$ are related by fusion with anyons in $\mathcal{A}_b$, and for every $\phi_i$ there is a unique anyon $a_i$ such that (\ref{braidingthm}) holds, we conclude that $\phi_i \bar{a}_i$ is independent of $i$. The parity flux in Eqs.~(\ref{factor1}) and (\ref{factor2}) is then simply defined as $\phi = \phi_i\bar{a}_i$ (a similar argument for the factorization of GAfTOs with fermion number symmetry was recently given in Ref. \cite{Lapa}).

As mentioned previously, the factorization of an AfTO $\mathcal{A}_f=\mathcal{A}_b\times\{1,f\}$ is not unique and we can obtain an equivalent factorization $\mathcal{A}_f=\mathcal{A}_b^\beta\times\{1,f\}$ using a homorphism $\beta:\mathcal{A}_b\rightarrow \mathbb{Z}_2$. So as consistency check, we show that for every $\mathcal{A}_b^\beta$, there exists a corresponding factorization of the GfTO as in Eqs.~(\ref{factor1}) and (\ref{factor2}). To see this, we can use the same lemma from Ref. \cite{Drinfeld} to conclude that for every homomorphism $\beta$, there must exist an anyon $\tilde{b}\in\mathcal{A}_b$ such that

\begin{equation}
(-1)^{\beta(a)}=M_{a,\tilde{b}}\,,\;\; \forall a \in\mathcal{A}_b
\end{equation}
So we see that now the GfTO factorizes as $\mathcal{A}_b^\beta\times \{1,\tilde{b}\phi ,f, \tilde{b}\phi f\}$ or $\mathcal{A}_b^\beta\times \{1,\tilde{b}\phi ,f\}$.

\subsubsection{Example: U$(1)_4\times$ $\overline{IQH}$}

Let us give an example to illustrate the factorization property of GAfTO. We consider the multi-component U$(1)$ Chern-Simons theory

\begin{equation}
\mathcal{L}=\frac{1}{4\pi}K_{IJ}\epsilon^{\mu\nu\lambda}a_\mu^I \partial_\nu a_\lambda^J + \frac{1}{2\pi}t_I \epsilon^{\mu\nu\lambda} A_\mu \partial_\nu a_{\lambda}^I\, ,
\end{equation}
with $K$-matrix
\begin{equation}
K = \left( \begin{array}{cc} 4 & \\ & -1 \end{array}\right)
\end{equation}
This describes a $\mathcal{A}_f=\mathcal{A}_b\times \{1,f\}=\mathbb{Z}_4\times \{1,f\}$ fermionic topological order, where the transparant fermion $f$ corresponds to the vector $l_f = \left(0,1\right)^T$, and $\mathbb{Z}_4=\{a,a^2,a^3,a^4=1\}$ is generated by anyon $a$ corresponding to vector $l_a=\left(1,0\right)^T$. The topological spins are given by $\theta_{a^p}=e^{ip^2\pi l_a^TK^{-1}l_a}=e^{ip^2\pi/4}$ and $\theta_f=e^{i\pi l_fK^{-1}l_f}=-1$.  $A_\mu$ is a probe gauge field for the global U$(1)$ particle number symmetry. If we require that the fermion has charge one, then this implies that $q_f=1=l_f^TK^{-1}t= -t_2$. So the only freedom left is the first component from the charge vector $t=(t_1, -1)^T$. This freedom determines the Hall conductance, which is given by $\sigma_{xy}=t^TK^{-1}t=t_1^2/4-1$, and the U$(1)$ charges of the anyons in $\mathbb{Z}_4$: $q_a = l_a^TK^{-1}t=t_1/4$. Let us take $t_1=1$, such that $\sigma_{xy}=-3/4$ and $q_a=1/4$. If a system has U$(1)$ fermion number symmetry, then there is a preferred way to create a parity flux by adiabatically inserting $\pi$ flux of the U$(1)$ particle number symmetry. Let us denote the fermion parity flux obtained via this adiabatic procedure as $\phi_A$. As is well-known, the topological spin of $\phi_A$ is fixed by the Hall conductance. In particular, it holds that $\theta_{\phi_A} = e^{i\pi\sigma_{xy}/4}$ \cite{Goldhaber,ChengZaletel}. If we apply this formula to our example with $t_1=1$, we learn that $\theta_{\phi_A}=e^{-i3\pi/16}$ and therefore $\theta_{\phi_A^2} = e^{-i3\pi/4}$. This implies that $\phi_A\times\phi _A= af$ or $\phi_A\times \phi_A = a^3f$, which is at odds with the proposed factorization property of the GAfTO because we cannot find a parity flux $\phi= \phi_A a^p$ such that $\phi\times\phi= f$. However, the choice $t_1=1$ is not allowed. This is because if $q_a=1/4$, then $q_{a^4}=q_1= 1$. This is not possible if $t$ is the charge vector of U$(1)$ fermion number symmetry, because the trivial anyon should always have even fermion parity. So $t_1=2t'$ has to be even. With this property correctly incorporated, we find $\theta_{\phi_A} = e^{i\pi (t'^2-1)/4}$ and $\theta_{\phi_A^2}=e^{i\pi(t'^2-1)}=(-1)^{t'+1}$. Under a shift $t'\rightarrow t'+4$, the anyon charges change as $q_a\rightarrow q_a + 2$, and $\theta_{\phi_A}$ remains invariant. So the only four remaining cases we have to consider are $t'=0,1,2,3$, corresponding to respectively $q_a = 0,1/2,1,3/2$. Let us work through the case where $t'=1$. With $t'=1$, it holds that $\phi_A\times\phi_A = a^2f$. We can now define $\phi = \phi_A a^3$, such that $\phi \times\phi = f$ and $M_{a,\phi}=M_{a,a^3}M_{a,\phi_A} = e^{i3\pi/2}e^{i\pi q_a} = 1$. Therefore, the GfTO factorizes as $\{a,a^2,a^3,1\}\times \{1,\phi_Aa^3,f,\phi_Aa^3f\}$. The factorization for other choices of $t'$ can be obtained in a similar way.

\section{Numerical identification of Abelian fermionic topological order}

An important question is how one can numerically identify the type of topological order realized by a particular microscopic lattice Hamiltonian. In the past decade, it has become clear that the ground state wavefunctions contain a lot of (if not all) information about the anyonic excitations. The first example of a ground-state property that can be used to diagnose topological order is the topological entanglement entropy  \cite{KitaevPreskill,LevinWen}, which gives access to the total quantum dimension. By looking not only at the entanglement entropy, but at the entire spectrum of the reduced density matrix corresponding to some spatial region in the ground state wavefunction one also obtains universal information about possible gapless edge modes of the system \cite{LiHaldane}. This correspondence between the entanglement spectrum and edge spectrum has been worked out in full detail for free fermion systems in Ref. \cite{Alexandradinata,Fidkowski}. For systems on a torus, Ref. \cite{ZhangGrover} identified the ground states with a definite anyon flux through one of the holes of the torus as those which are `minimally entangled states' (MES) with respect to cuts wrapping the hole under consideration. In the MES basis, one can obtain both the $S$ and $T$ matrices by taking certain wave function overlaps \cite{ZhangGrover,ZaletelMong}. Finally, using the same MES on the cylinder, one can also find the $T$ matrix by calculating the `momentum polarization' \cite{ZaletelMong,TuZhang,Cincio,Zaletel,He,Wen}. 

In section \ref{sec:mompol}, we first discuss the application of the momentum polarization technique to systems with non-trivial translational symmetry fractionalization, characterized by the presence of a non-trivial background anyon in each unit cell. Next to the entanglement contribution to momentum polarization as discussed in Refs. \cite{ZaletelMong,TuZhang,Cincio,Zaletel,He,Wen}, we identify an `eigenvalue contribution' which is determined by both the background anyon and the anyonic flux which labels the MES. The interplay of the entanglement and eigenvalue contributions to the momentum polarization is shown to lead to a physically intuitive picture for the behavior of the momentum polarization under a change of entanglement cut, from which one can numerically obtain the topological spin of the background anyon. Next, we discuss the application of the momentum polarization technique to fermion systems. It is shown that in fermionic systems not only non-trivial background anyons in $\mathcal{A}_b$, but also transparant `background fermions' give rise to a non-trivial eigenvalue contribution to the momentum polarization, which results entirely from the fermionic anti-commutation relations of the microscopic fermions. This eigenvalue contribution from background fermions also fits nicely with the physically intuitive picture for the dependence of the momentum polarization on the entanglement cut. 

In section \ref{sec:numid} we use the results of section \ref{sec:gauging} to describe a minimal scheme to numerically identify a general AfTO in numerics, without relying on charge conservation symmetry. In particular, we make use of the special structure of GAfTO's to show that, compared to bosonic systems, the only additional piece of information that needs to be determined is $\theta_\phi$, i.e. the topological spin of the fermion parity flux. We provide a concrete (minimal) procedure to obtain $\theta_\phi$ by calculating the momentum polarization, for which one can rely on the results of section \ref{sec:mompol}.

\subsection{Momentum polarization in the presence of background anyons and its application to fermion systems}\label{sec:mompol}

\subsubsection{Bosonic systems}

Before turning to fermion systems, we first discuss the concept of momentum polarization in boson or spin systems,. Consider a MES on a cylinder with an anyon flux of the type $a$ through the hole of the cylinder. Let us denote this MES as $|\psi[a]\rangle$. We will call the direction along the axis of the cylinder the $x$-direction, and the direction wrapping the hole the $y$-direction. The size of the cylinder is given by $N_x\times N_y$ unit cells. We now choose a cut along the $y$-direction close to the middle of the cylinder, dividing the cylinder in two. The length of the left half is then $N_x^L$, while the length of the right half is $N_x^R$, such that $N_x^L+N_x^R=N_x$. With this cut, the translation operator in the $y$-direction can be written as a tensor product between the translation operator on the left half and the translation operator on the right half: $T_y = T_y^L\otimes T_y^R$. With these definitions in place, momentum polarization was defined in Refs. \cite{ZaletelMong,TuZhang,Cincio} as the expectation value $\langle\psi[a]|T_y^L|\psi[a]\rangle$. It was found that this expectation value scales with $N_y$ as

\begin{equation}\label{mompol}
\langle\psi[a]|T_y^L|\psi[a]\rangle = \exp\left(\frac{2\pi i}{N_y}\left( h_a -\frac{c_-}{24}\right) - \alpha N_y \right)\, ,
\end{equation}
where $\theta_a = e^{2\pi i h_a}$ is the topological spin of anyon $a$, and $c_-$ is the chiral central charge. The complex number $\alpha$ is non-universal. 

One important assumption for the validity of Eq. (\ref{mompol}) is that there is no translational symmetry fractionalization. As we explain in more detail below, with non-trivial translational symmetry fractionalization we find that Eq. (\ref{mompol}) needs to be generalized to the following more general form:

\begin{equation}\label{mompolgeneral}
\langle\psi_b[C,a]|T_y^{C,L}|\psi_b[C,a]\rangle = \exp\left(\frac{2\pi i}{N_y}\left( h_a -\frac{c_-}{24}\right) + i\Theta_{a,b}N_x^{C,L} - \alpha_C N_y \right)\, ,
\end{equation}
where the notation for the MES $|\psi_b[C,a]\rangle$ now depends on two anyon labels $a$ and $b$, and on an integer $C$ which denotes the position of the entanglement cut. The anyon $a$ is again the anyon flux through the hole of the cylinder, measured along cut $C$, and $b$ is a `background anyon' which sits inside every unit cell \cite{Zaletel3,ChengZaletel}. The translation operator $T^{C,L}_y$ acts on the left of the cut labeled by $C$. On the right hands side, $N_x^{C,L}$ is an integer which corresponds to the length of the left half of the cylinder, and $\alpha_C$ is a non-universal complex number also depending on the cut. Although $\alpha_C$ is non-universal, we will argue that the difference $\Delta\alpha_C = \alpha_{C+1}-\alpha_C$ between two neighboring cuts is universal. The interpretation of $\Theta_{a,b}$ will be explained in the next paragraph. 

To explain the general form of the momentum polarization formula Eq. (\ref{mompolgeneral}), it is useful to decompose the momentum polarization in an `eigenvalue contribution' and an `entanglement contribution'. Let us consider translationally invariant systems, such that $|\psi_b[C,a]\rangle$ is an eigenstate of $T_y$. The $y$-momentum of the MES is then determined by the anyon flux $a$ and the background anyon $b$ as follows:

\begin{equation}\label{Teig}
T_y|\psi_b[C,a]\rangle = e^{ i \Theta_{a,b} N_x}|\psi_a \rangle\, ,
\end{equation}
where $e^{i\Theta_{a,b}} = M_{a,b}$  \cite{Zaletel3,ChengZaletel}. This background anyon has to be non-trivial in systems where a Lieb-Schultz-Mattis-Oshikawa-Hastings (LSMOH) obstruction to a gapped trivial featureless phase is present \cite{LSM,Oshikawa1,Hastings,Oshikawa2,Zaletel3}. Equation (\ref{Teig}) holds irrespective of whether the system is bosonic or fermionic, although in general the types of topological orders which can satisfy the LSMOH obstruction are different in both cases \cite{FillingC}. From Eq. (\ref{Teig}) we can immediately identify the eigenvalue contribution to the momentum polarization as $e^{i\Theta_{a,b}N_x^{C,L}}$.

To identify the entanglement contribution to momentum polarization (as discussed in Refs. \cite{ZaletelMong,TuZhang,Cincio,Zaletel,He,Wen}), we write the MES as

\begin{eqnarray}
|\psi_b[C,a]\rangle & = & \sum_{\alpha,\beta} \Psi_{C,a,b}^{\alpha,\beta}|\alpha\rangle_L\otimes|\beta\rangle_R \\
 & = & \sum_{\mu} s_{C,a,b}^\mu\, |\mu\rangle_L\otimes |\mu\rangle_R
\end{eqnarray}
In the first line, we have used an arbitrary basis $|\alpha\rangle_L$ ($|\beta\rangle_R$) for the left (right) half of the cylinder. In the second line, the state is decomposed in the Schmidt basis. In the Schmidt basis, the action of translation in the $y$-direction can be written as

\begin{eqnarray}
T_y|\psi_b[C,a]\rangle & = & \sum_\mu s_{C,a,b}^\mu\,  T^{C,L}_y|\mu\rangle_L\otimes T_y^{C,R}|\mu\rangle_R \\
 & = &  e^{ i \Theta_{a,b} N_x}\sum_{\mu\lambda\sigma} s_{C,a,b}^\mu [U^{*C,L}_{a,b}]_{\mu\lambda} [U^{C,R}_{a,b}]_{\mu\sigma}|\lambda\rangle_L\otimes|\sigma\rangle_R\, ,
\end{eqnarray}
where in the second line, we have re-expanded $T^{C,L/R}_y|\mu\rangle_{L/R}$ in the Schmidt basis. Note that since $T_y^{C,L}$ and $T_y^{C,R}$ are unitary, so are $U^{C,L}_{a,b}$ and $U^{C,R}_{a,b}$. We have also separated out the eigenvalue factor $e^{i\Theta_{a,b}N_x}$. The MES $|\psi_b[C,a]\rangle$ is an eigenstate of $T_y$ if

\begin{equation}
\left(U_{a,b}^{C,L}\right)^\dagger S_{C,a,b}U_{a,b}^{C,R} = S_{C,a,b}\, ,
\end{equation}
where $S_{C,a,b} =$diag$(s_{C,a,b}^\mu)$. This implies that $U^{C,L}_{a,b}=U^{C,R}_{a,b}=U^C_{a,b}$, where $U^C_{a,b}$ commutes with $S_{C,a,b}$. The entanglement contribution to momentum polarization is then entirely given in terms of the Schmidt values and the unitary matrix $U^C_{a,b}$, and takes the form

\begin{equation}
\mathrm{tr}(S^2_{C,a,b}U^{C}_{a,b}) = \exp\left(\frac{2\pi i}{N_y}\left( h_a -\frac{c_-}{24}\right) - \alpha_C N_y \right) \, ,
\end{equation}
where we have again used the notation $\alpha_C$ to emphasize that $\alpha_C$ depends on the choice of cut. To understand how $\alpha_C$ depends on the cut, we note that because of the background anyon $b$ per unit cell, the anyon flux through the hole of the cylinder is not the same for every cut. The anyon fluxes for two neighboring cuts differ by the total background anyon charge enclosed by the two cuts, which is $b^{N_y}$. In other words, it holds that

\begin{equation}
|\psi_b[C,a]\rangle = |\psi_b[C+1,ab^{N_y}]\rangle
\end{equation}
The ribbon identity allows us to write $\theta_{ab^{Ny}} = \theta_a\theta_{b}^{N_y^2}M_{a,b}^{N_y}$. So given the momentum polarization for a particular cut, we can obtain the momentum polarization for the neighboring cut by shifting

\begin{equation}\label{intuitive}
h_a \rightarrow h_a + h_b N_y^2 + \frac{\Theta_{a,b}}{2\pi}N_y
\end{equation}
From this we conclude that $\alpha_C$ depends on the cut as $\alpha_{C+1}=\alpha_C +2\pi i h_b$. So, interestingly, even though $\alpha_C$ is non-universal, by calculating this coefficient for two neighboring cuts one can numerically obtain the topological spin of the background anyon $b$. Also, note that Eq. (\ref{intuitive}) implies that even though the anyon flux through the hole of the cylinder depends on the choice of entanglement cut if $b$ is non-trivial, the momentum polarization nevertheless allows one to obtain a topological spin $h_a$ which is independent of the choice of cut, because it relies on a scaling in the cylinder circumference $N_y$.

\subsubsection{Fermionic systems}

For fermion systems on a cylinder, there are two types of boundary conditions: periodic and anti-periodic. For each type of boundary condition, one can find a set of MES. Let us start by considering MES in the anti-periodic sector.

\emph{Anti-periodic boundary conditions --}  With anti-periodic boundary conditions, the MES have anyon fluxes through the hole of the cylinder which are labeled by the anyons in $\mathcal{A}_f$, and \emph{not} by the different types of parity fluxes. The latter label MES in the \emph{periodic} sector, which we discuss below. So let us write a MES in the anti-periodic sector as

\begin{equation}
|\psi_{f^\sigma b}[C,f^\lambda a]\rangle\, , \;\;\;\sigma,\lambda\in \{0,1\}\,, \;\;a,b \in \mathcal{A}_b\, ,
\end{equation}
where $f^\sigma b$ again denotes the background anyon and $f^\lambda a$ the anyon flux through the hole of the cylinder measured at cut $C$. Note that both the background anyon and the flux through the hole of the cylinder are labeled with anyons in $\mathcal{A}_f=\mathcal{A}_b\times\{1,f\}$, even though the ground state degeneracy on the torus is only given by $|\mathcal{A}_b|$, i.e. the number of anyons in $\mathcal{A}_b$. 

In the anti-periodic sector, the translation operator along the $y$-direction is defined as

\begin{eqnarray}
\tilde{T}_y: & c^\dagger_{(x,y)}\rightarrow c^\dagger_{(x,y+1)}\,,\qquad y\neq N_y-1\\
 & c^\dagger_{(x,N_y-1)}\rightarrow -c^\dagger_{(x,1)}
\end{eqnarray}
Using this twisted translation operator, the momentum polarization is defined as

\begin{equation}\label{mompolAP}
\langle\psi_{f^\sigma b}[C,f^\lambda a]|\tilde{T}^{C,L}_y|\psi_{f^\sigma b}[C,f^\lambda a] \rangle = \exp\left(\frac{2\pi i}{N_y}\left( h_{f^\lambda a} -\frac{c_-}{24}\right) + i\Theta_{a,b}N_x^{C,L} - \alpha_C N_y \right)
\end{equation}
As in bosonic systems, the MES satisfy the following property:

\begin{equation}
|\psi_{f^\sigma b}[C,f^\lambda a]\rangle = |\psi_{f^\sigma b}[C+1,f^{\lambda+\beta N_y} ab^{N_y}]\rangle\, ,
\end{equation}
which via the replacement $h_{f^\lambda a}\rightarrow h_{f^\lambda a} + N_y^2 (\sigma h_f +h_b) + \Theta_{a,b}N_y/2\pi$ implies that $\alpha_{C+1}-\alpha_C = 2\pi i (\sigma h_f +h_b) = i(\sigma \pi + 2\pi h_b)$. The dependence of $\Delta\alpha_C$ on $\sigma$ arises from the eigenvalue contribution to the entanglement polarization. To see this, consider the situation where $\mathcal{A}_b=1$. In this case, $\sigma$ corresponds to the fermion parity per unit cell, which means that  on the torus the fermion parity of $|\psi_{f^\sigma}[C,1]\rangle$ is given by $(-1)^{\sigma N_x N_y}$. The momentum in the $y$-direction on the torus (with periodic boundary conditions in the $x$-direction) is then given by

\begin{equation}\label{transeig1}
\tilde{T}_y|\psi_{f^\sigma}[C,1]\rangle = (-1)^{\sigma N_x N_y}|\psi_{f^\sigma}[C,1]\rangle
\end{equation}
This property can readily be checked for band insulators, where $\sigma$ is the number of filled bands modulo $2$, and also follows from fermionic tensor network descriptions of gapped ground states \cite{fMPS,fPEPS}. From Eq. (\ref{transeig1}) we can identify the eigenvalue contribution to the momentum polarization as $e^{i\sigma\pi N_x^{C,L}N_y}$, which indeed leads to the dependence of $\alpha_C$ on the cut as described above.

The last aspect of the MES in the anti-periodic sector that we need to comment on is the role of $f^\lambda$. The choice of $\lambda = 0,1$ is a non-universal property of the MES, as can be seen by noting that the value of $\lambda$ can be flipped by adding a single electron in a $k_y = \pi/N_y$ momentum state on the left of the cut, which changes the (eigenvalue contribution to the) momentum polarization accordingly by a factor of $e^{i\pi/N_y}$. This reflects the fact that the ground state degeneracy on the torus is given by $|\mathcal{A}_b|$, and not $|\mathcal{A}_f|$.

\emph{Periodic boundary conditions --} On a cylinder with periodic boundary conditions, the anyonic fluxes which label the MES are given by the different fermion parity fluxes. So, with anti-periodic boundary conditions, we can write the MES as

\begin{equation}
|\psi_{f^\sigma b}[C,\phi a f^\lambda]\rangle\, , \;\;\;\sigma,\lambda\in \{0,1\}\,, \;\;a,b \in \mathcal{A}_b\, ,
\end{equation}
where, as before, $\phi$ is the fermion parity flux which braids trivially with all anyons in $\mathcal{A}_b$. The momentum polarization in the periodic sector is then given by

\begin{equation}\label{mompolP}
\langle\psi_{f^\sigma b}[C,\phi a f^\lambda]| T_y^{C,L}|\psi_{f^\sigma b}[C,\phi a f^\lambda]\rangle = \exp\left(\frac{2\pi i}{N_y}\left( h_{\phi a} -\frac{c_-}{24}\right) + i\Theta_{\phi a,bf^\sigma}N_x^{C,L} - \alpha_C N_y \right)
\end{equation}
The by now familiar property of the MES:

\begin{equation}
|\psi_{f^\sigma b}[C,\phi a f^\lambda]\rangle = |\psi_{f^\sigma b}[C+1,\phi ab^{N_y} f^{\lambda+\sigma N_y}]\rangle
\end{equation}
implies that $\alpha_C$ depends on the cut as $\alpha_{C+1}-\alpha_C = i(\sigma \pi + 2\pi h_b)$. As a consistency check, let us again take $\mathcal{A}_b = 1$ such that $\sigma$ is the fermion number per site. This means that if we define the state on the torus with periodic boundary conditions along both cycles, then it holds that

\begin{equation}
(-1)^{\hat{F}}|\psi_{f^\sigma }[C,\phi ]\rangle = (-1)^{\eta+\sigma N_x N_y}|\psi_{f^\sigma }[C,\phi ]\rangle \, ,
\end{equation}
where $\hat{F}$ is the fermion number operator, and $\eta=1$ if $\phi$ is non-Abelian \cite{ReadGreen} and $\eta=0$ otherwise. With this definition, one finds that on a torus the momentum in the $y$-direction is given by 

\begin{equation}\label{neighboringcuts}
T_y|\psi_{f^\sigma }[C,\phi ]\rangle = (-1)^{ \sigma N_x(N_y + 1)} |\psi_{f^\sigma }[C,\phi ]\rangle\, ,
\end{equation}
Again, this property can easily be checked for band insulators, and can also be seen in the fermionic tensor network formalism \cite{fMPS,fPEPS}. Eq. (\ref{neighboringcuts}) implies that the momentum polarizations for two neighboring cuts indeed differ by a factor $(-1)^{\sigma(N_y +1)} = e^{i\left(\Theta_{\phi,f^\sigma}+\sigma \pi N_y\right)}$, which arises entirely from the eigenvalue contribution to the momentum polarization.

Finally, we note that the momentum polarization with periodic boundary conditions is independent of $f^\lambda$. This agrees with the fact that the topological spins of the fermion parity fluxes are invariant under the addition of a transparent fermion.

In the appendix, we illustrate our general discussion of momentum polarization in fermionic systems by applying it to a Chern insulator and a topological $p+ip$ superconductor. In particular, we show that both in the anti-periodic and periodic sectors, the dependence of the momentum polarization on the choice of cut is indeed captured respectively by Eqs. (\ref{mompolAP}) and (\ref{mompolP}), with $\alpha_{C+1}-\alpha_C = i\pi$.

\subsection{Numerically determining the topological order}\label{sec:numid}

Having uncovered the special structure of GAfTO's in section \ref {sec:gauging}, and the details of momentum polarization in fermionic systems in section \ref{sec:mompol}, we now have all the necessary ingredients at our disposal to outline a minimal numerical detection scheme for the most general AfTO. First, by calculating the momentum polarizations for the different MES on the cylinder in the anti-periodic sector we obtain the topological spins of the anyons in $\mathcal{A}_b$, up to a minus sign ambiguity. Secondly, for the MES on the torus with anti-periodic boundary conditions along both cycles, we can use the formalism of Refs. \cite{ZaletelMong,ZhangGrover,Cincio} to obtain the unitary $S$-matrix of the MTC corresponding to $\mathcal{A}_b$, just as one does for bosonic systems. From the $S$-matrix, one obtains the fusion rules (i.e. the group structure) of $\mathcal{A}_b$ via the Verlinde formula \cite{Verlinde}

\begin{equation}
N_{a,b}^c = \sum_d \frac{S_{a,d}S_{b,d}S_{d,c}^*}{S_{0,d}}\, ,
\end{equation}
Once the group structure of $\mathcal{A}_b$ is obtained, this can be used to partially fix the sign ambiguity in the topological spins of the anyons in $\mathcal{A}_b$, by imposing that $S_{a,b} = N^{-1/2}M_{a,b} = N^{-1/2}\theta_{ab}/(\theta_a\theta_b)$, where $N$ is the number of MES (= the number of anyons in $\mathcal{A}_b$). This fixes the topological spins up to a homomorphism $\beta$ from $\mathcal{A}_b$ to $\mathbb{Z}_2$. This is the same homomorphism as discussed above, and different $\beta$ correspond to different ways of writing $\mathcal{A}_f = \mathcal{A}_b^\beta\times \{1,f\}$, where $\mathcal{A}_b^\beta = \{af^{\beta(a)}|a\in\mathcal{A}_b\}$. At this point, one has to choose a particular homomorphism to fix the topological spins.

To completely determine the topological quantum phase of the system of interest one does not only need to know $\mathcal{A}_f$, but also the complete algebraic data corresponding to the modular extension $\mathcal{A}_b\times \{1,f,\phi,f\phi\}$ (if $\phi$ is Abelian) or $\mathcal{A}_b\times\{1,f,\phi\}$ (if $\phi$ is non-Abelian). Let us first consider the case where the fermion parity fluxes are Abelian. Since $\phi$ is Abelian, it holds that either $\phi \times \phi = 1$ or $\phi\times\phi=f$. This in turn implies that $\theta_\phi^4=1$ or $\theta_\phi^4=-1$, which gives us eight different possible values for $\theta_\phi$. Once we know the topological spin of the Abelian parity flux, we have completely fixed which of the eight possible modular extensions is realized. When $\mathcal{A}_b=1$, this was shown by Kitaev as a part of his `$16$-fold way' \cite{Kitaev}.

A non-Abelian parity flux satisfies $\phi\times\phi=1+f$. If $\mathcal{A}_b=1$, then Kitaev has shown that there exist exactly eight different modular extensions with non-Abelian parity fluxes \cite{Kitaev}. These eight different modular extensions correspond to the eight different Ising MTC's, and are uniquely identified by the topological spin of the fermion parity flux. In Ref. \cite{Wang}, it was conjectured that Kitaev's 16-fold way generalizes to all fTO's (Abelian and non-Abelian), i.e. it was conjectured that every fTO has exactly 16 different modular extensions, 8 of which have Abelian fermion parity flux and 8 which have non-Abelian fermion parity flux. When the fTO is Abelian, it was shown above that the GfTO factorizes as $\mathcal{A}_b\times\{1,f,\phi\}$, so there are indeed eight different non-Abelian modular extensions which again correspond the eight different Ising categories.

From the above discussion, we learn that to complete the numerical identification of an Abelian fermionic topological order we only need to know $\theta_\phi$. To access $\theta_\phi$, one first calculates the momentum polarizations of the MES on the cylinder with periodic boundary conditions. This provides a set of topological spins, which correspond to $\theta_{a\phi}$. Because $\phi$ braids trivially with all anyons in $\mathcal{A}_b$, these topological spins factorize as $\theta_{a\phi} = \theta_a \theta_\phi$. This implies that we can organize the topological spins in the periodic sector in an ordered vector, which is proportional to the ordered vector of topological spins in the anti-periodic sector. The proportionality constant obtained in this way is unique and corresponds to $\theta_\phi$, such that one can simply try all the possible permutations of the topological spins in the periodic sector until one finds one where the required proportionality is realized. To see why the proportionality constant is unique we need to show that we cannot permute the vector of topological spins in the periodic sector to obtain a vector that is proportional to (and different from) the original, unpermuted one. Since we are only interested in permutations that do not leave the vector invariant, there can be no element in the vector that is fixed under the permutation. This means that the permutation acts as $\theta_{a\phi} \rightarrow \theta_{ad\phi}$, where $d$ is an anyon in $\mathcal{A}_b$ which is not the trivial anyon. Because $\theta_{ad\phi} = \theta_{a\phi}\theta_d M_{a,d}$, the resulting vector is proportional to the unpermuted one if and only if $M_{a,d}$ is independent of $a$. But this is a violation of modularity, and therefore this cannot happen. This completes the procedure of how to uniquely characterize an AfTO in numerics.

Before concluding the discussion on numerical identification of fermionic topological orders, let us consider what happens if one would have made a different choice of homomorphism $\beta$. In that case, $\mathcal{A}_b$ becomes $\mathcal{A}_b^\beta = \{af^{\beta(a)}|a\in \mathcal{A}_b\}$ and one would interpret the set of topological spins obtained from momentum polarization in the periodic sector as $\theta_{af^{\beta(a)}\tilde{\phi}} = \theta_{af^{\beta(a)}\phi \tilde{b}}$, where $\tilde{b}$ is the unique anyon that satisfies $(-1)^{\beta(a)} = M_{a,\tilde{b}}$. Using this property, one can factorize the topological spins in the periodic sector as

\begin{equation}
\theta_{af^{\beta(a)}\tilde{\phi}}=\theta_{af^{\beta(a)}\phi \tilde{b}} = \theta_{af^{\beta(a)}}\theta_{\phi \tilde{b}} = \theta_{af^{\beta(a)}}\theta_{\tilde{\phi}} \, ,
\end{equation}
so we can again permute them in such a way that they become proportional --as a vector-- to the vector of topological spins $\theta_{af^{\beta(a)}}$ in the anti-periodic sector. Of course, the set of topological spins in the periodic sector obtained from momentum polarization is independent of our choice of $\beta$, as can easily be verified:

\begin{eqnarray}
\theta_{af^{\beta(a)}\tilde{\phi}} = \theta_{af^{\beta(a)}\phi \tilde{b}}=  \theta_{a\tilde{b}\phi}
\end{eqnarray}
So we find that the final identification of the AfTO is independent of our choice of $\beta$, as it should be of course. One only needs to keep in mind that when comparing two different AfTO with the same $\mathcal{A}_f$, one should always use the same of choice of $\beta$ to compare the topological spins of the fermion parity fluxes $\phi$.

\section{Gauging fermion parity with boundaries}

\subsection{Fermionic vs. bosonic gapped edges}

\begin{figure}
\begin{center}
\includegraphics[scale=0.5]{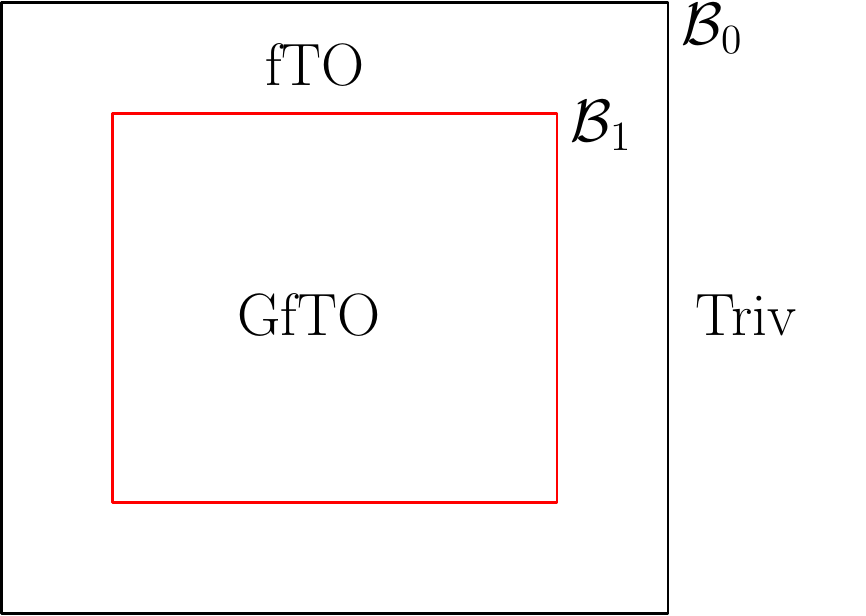}
\caption{A GfTO obtained by gauging a fTO with a gapped boundary $\mathcal{B}_0$ to the trivial phase. The gauging is done such that a narrow strip along the boundary is unaffected and remains in the original fTO phase. The boundary separating the GfTO from the ungauged fTO is denoted as $\mathcal{B}_1$.}\label{fig:B1}
\end{center}
\end{figure}

In this section, we address the question of what happens at the boundary of an AfTO after fermion parity is gauged. To gain some intuition about this question, let us consider a fTO which has a gapped boundary $\mathcal{B}_0$ separating it from the trivial phase. Now imagine gauging the fermion parity everywhere in the bulk, except in a narrow strip along the edge. After gauging, the GfTO in the bulk is separated by a boundary $\mathcal{B}_1$ from a narrow strip of the original ungauged fTO, which itself is separated from the trivial phase by the gapped boundary $\mathcal{B}_0$. See figure \ref{fig:B1}(a) for an illustration. If $\mathcal{B}_1$ is also gapped, then this construction gives a gapped boundary separating the GfTO from the trivial phase.

To argue why we can always take $\mathcal{B}_1$ to be gapped, let us first consider the case of bTO. In bosonic systems, the `ungauging' procedure corresponds to condensing the gauge charges \cite{Barkeshli}, which are always bosonic and braid trivially with each other. Condensing the gauge charges results in confinement of the gauge fluxes, which means that after condensation the energy of a flux pair grows linearly with the spatial separation between the fluxes. Condensation of the gauge charges thus transforms the gauge fluxes into the symmetry defects of the ungauged phase \cite{Barkeshli}. Using the general relation between anyon condensation and gapped boundaries \cite{Levin,KitaevKong}, we can then always construct the gapped boundary $\mathcal{B}_1$ between the gauged and ungauged bTO as a domain wall where the gauge charges get condensed. 

In GfTOs, the $\mathbb{Z}_2$ gauge charge $\tilde{f}$ is by definition a fermion. Because $\tilde{f}$ has non-trivial topological spin, it is impossible to construct a gapped boundary where $\tilde{f}$ gets condensed. However, at a domain wall between the GfTO and the original fTO, we can condense the bound state $\tilde{f}f$, where $f$ is the transparant fermion of the fTO. Because $\tilde{f}f$ is a boson, this will result in a gapped boundary. See Refs. \cite{Aasen,Wan} for more details on this construction. 

At this point, we have obtained an argument that gauging fermion parity always preserves a gapped edge. Let us now connect this argument to the formalism of Lagrangian subgroups reviewed above in Secs. \ref{GappedB} and \ref{GappedF}, where it was stated that an Abelian bosonic (fermionic) TO admits a gapped edge iff it has a bosonic (fermionic) Lagrangian subgroup. Since the GfTO is modular it corresponds to a bTO, and so according to Sec. \ref{GappedB} we would expect it to have a bosonic Lagrangian subgroup if it admits a gapped boundary. However, this is far from clear from the argument presented above. In the `layered-boundary' construction the fermionic bulk gauge charge $\tilde{f}$ is bound to a transparant, microscopic fermion $f$  which only lives on the edge and the resulting bound state is subsequently condensed. We will refer to such a gapped edge which relies on the presence of microscopic boundary fermions as a `fermionic gapped edge'. The conventional notion of a gapped edge for a bTO as used in Sec. \ref{GappedB}, however, does not permit the use of microscopic fermionic degrees on the boundary, and requires all condensed particles to be bosons. We will refer to such a gapped edge which does not contain microscopic fermion degrees of freedom as a `bosonic gapped edge'. Using this terminology, Secs. \ref{GappedB} and \ref{GappedF} then simply state that an Abelian bosonic (fermionic) TO admits a bosonic (fermionic) gapped edge iff it has a bosonic (fermionic) Lagrangian subgroup. So we see that the `layered-boundary' argument only tells us that we can construct a fermionic gapped edge for the GfTO, but not necessarily a bosonic gapped edge. 

In the next two sections, we will show the stronger statement that the GfTO always admits a bosonic gapped edge if the ungauged fTO admits a fermionic gapped edge. We will also show the inverse implication, i.e. that an AfTO has a fermionic gapped edge if the gauged theory has a bosonic gapped edge. One of the implications of this result is that any bosonic topological order which is obtained by gauging a fermionic topological order with a gapped edge, can be realized by a purely bosonic lattice Hamiltonian with a gapped edge. 

\subsection{Abelian parity flux}

In this section, we will show that a GAfTO has a bosonic gapped edge iff the corresponding ungauged AfTO has a fermionic gapped edge, while assuming that the fermion parity flux is Abelian. The case with non-Abelian fermion parity flux will be discussed in the next section.

When the parity fluxes are Abelian it is not difficult to see that if we apply Eq.~(\ref{kit}) to the modular GAfTO, we get the following expression:

\begin{equation}\label{GM1}
e^{2\pi ic_-/8} = \frac{\theta_\phi}{\sqrt{N_b}}\sum_{a\in\mathcal{A}_b}\theta_a \,,
\end{equation}
where $N_b$ is the number of anyons in $\mathcal{A}_b$. Because gauging a discrete symmetry cannot change the chiral central charge, Eq. (\ref{GM1}) not only determines $c_-$ of the GAfTO (mod 8), but also of the original ungauged fTO \cite{TianLan1,TianLan2}. 

If we apply Eq. (\ref{GM1}) to a different factorization $\mathcal{A}_b^\beta\times\{1,\tilde{b}\phi,f,\tilde{b}\phi f\}$ of the GAfTO, we get

\begin{eqnarray}
e^{2\pi ic_-/8} & = &  \frac{\theta_{\tilde{b}\phi}}{\sqrt{N_b}}\sum_{a\in\mathcal{A}_b^\beta}\theta_a \\
 & = & \frac{\theta_{\phi}\theta_{\tilde{b}}}{\sqrt{N_b}} \sum_{a\in\mathcal{A}_b}\theta_a (-1)^{\beta(a)} \label{GM2}
\end{eqnarray}
Because $M_{a,\tilde{b}}=(-1)^{\beta(a)}$ for all $a\in\mathcal{A}_b$, it follows that $\tilde{b}^2$ braids trivially with all anyons in $\mathcal{A}_b$, and must therefore be the trivial anyon. From $\tilde{b}^2=1$, we know that $\theta_{\tilde{b}}^4=1$. So by equating Eqs. (\ref{GM1}) and (\ref{GM2}), we find that the insertion of the minus signs $(-1)^{\beta(a)}$ in the sum of the topological spins of $\mathcal{A}_b$ changes the value of that sum by a multiplicative factor which is a fourth root of unity, and equals $\theta_{\tilde{b}}^*$. Now we are equipped to show the following result: 
\begin{center}
\emph{A GAfTO has a bosonic Lagrangian subgroup $\mathcal{L}_b$ if and only if the corresponding ungauged AfTO has a fermionic Lagrangian subgroup $\mathcal{L}_f$, and an integer chiral central charge which is a multiple of eight.} 
\end{center}
Using the bulk-boundary connection reviewed in the previous sections, this result then implies that a GfTO has a bosonic gapped edge if and only if the ungauged fTO has a fermionic gapped edge.

We first show the `if' direction and assume that the fTO has a fermionic Lagrangian subgroup $\mathcal{L}_f$ and zero chiral central charge (mod 8). To start, we observe that because the anyons in $\mathcal{L}_f$ braid trivially with each other, it follows from the ribbon identity that their topological spins form a $\mathbb{Z}_2$ valued representation of $\mathcal{L}_f$. In other words, $\theta_{l_i}=(-1)^{\alpha(l_i)}$ such that $\alpha(\cdot):\mathcal{L}_f\rightarrow \mathbb{Z}_2=\{0,1\}$ is a homomorphism. In general this homomorphism cannot be extended to a homomorphism from $\mathcal{A}_b$ to $\mathbb{Z}_2$. Using this homomorphism, we define $\mathcal{L}'_f = \{l_if^{\alpha(l_i)}|l_i\in\mathcal{L}_f\}$ such that all anyons in $\mathcal{L}_f'$ have trivial topological spin. We now want to extend $\mathcal{L}'_f$ in such a way that it becomes a bosonic Lagrangian subgroup of the GAfTO. Because all anyons in $\mathcal{L}_f'$ braid trivially with $f$ and because $f$ itself cannot be in $\mathcal{L}_b$, the extended Lagrangian subgroup $\mathcal{L}_b$ will have to contain a fermion parity flux. At this point, it is clear that we can find a bosonic Lagrangian subgroup $\mathcal{L}_b=\mathcal{L}'_f\times\{1,\phi c\}$ of the GfTO, provided that there exists an anyon $c\in \mathcal{A}_b$ such that: (1) $M_{c,l}= (-1)^{\alpha(l)}$ for all $l\in\mathcal{L}_f$, and (2) $\theta_\phi = \theta_c^*$. Here, we have used the factorization property of GAfTO to define $\phi$ as the parity flux which braids trivially with all anyons in $\mathcal{A}_b$.

Let us first show that there exists an anyon $c$ satifying property (1), i.e. $M_{c,l}=(-1)^{\alpha(l)}$ for all $l\in\mathcal{L}_f$. If the homomorphism $\alpha: \mathcal{L}_f\rightarrow \mathbb{Z}_2$ is trivial, then property (1) is also trivial and $c$ is simply the identity anyon. Let us therefore focus on the case where $\alpha$ is non-trivial and write $\mathcal{A}_b = \{\mathcal{L}_f, \mathcal{L}_f\times c_1, \mathcal{L}_f\times c_2,\dots,\mathcal{L}_f\times d_1,\mathcal{L}_f\times d_2,\dots \}$, where $c_i, d_i$ are a set of anyons not in $\mathcal{L}_f$ of which the $c_i$ satisfy criterion (1), and the $d_i$ do not. Using Eq. (\ref{GM1}), we find

\begin{eqnarray}
\sqrt{N_b}\theta_\phi^* e^{2\pi ic_-/8} & = & \sum_{l\in\mathcal{L}_f}\theta_l + \sum_{i}\sum_{l\in\mathcal{L}_f}\theta_{lc_i}+\sum_{i}\sum_{l\in\mathcal{L}_f}\theta_{ld_i} \\
& = & \sum_{i}\theta_{c_i}\sum_{l\in\mathcal{L}_f}(-1)^{\alpha(l)}M_{l,c_i}+\sum_{i}\theta_{d_i}\sum_{l\in\mathcal{L}_f}(-1)^{\alpha(l)}M_{l,d_i} \\
& = & N_{L_f}\sum_{i}\theta_{c_i}\label{ci}\, ,
\end{eqnarray}
where $N_{L_f}$ is the number of anyons in $\mathcal{L}_f$. In the second and third line we have applied Schur's orthogonality relations to the 1D irreps of $\mathcal{L}_f$. From this result, we see that there must exist at least one anyon $c_i$. 

To show that we can find an anyon $c$ satisfying both properties $(1)$ ($M_{c,l}=(-1)^{\alpha(l)}$), and $(2)$ ($\theta_c=\theta_\phi^*$), we proceed as follows. Because the $c_ic_j$ braid trivially with all anyons in $\mathcal{L}_f$, it follows from the definition of a fermionic Lagrangian subgroup that $c_ic_j \in \mathcal{L}_f$. This implies that

\begin{equation}
\theta_{c_ic_j} = (-1)^{\alpha(c_ic_j)} =  M_{c_i,c_ic_j} 
\end{equation}
Applying the ribbon identity to the left-hand side of this equation, we find

\begin{equation}
\theta_{c_i}\theta_{c_j} M_{c_i,c_j} =  M_{c_i,c_i}M_{c_i,c_j} = \theta_{c_i}^2 M_{c_i,c_j} \Rightarrow \theta_{c_i} = \theta_{c_j}
\end{equation}
Because the topological spins of all the $c_i$ are the same, expression (\ref{ci}) for the chiral central charge becomes

\begin{equation}
e^{2\pi ic_-/8} = \frac{N_{L_f}}{\sqrt{N_b}}N_c \theta_\phi \theta_c \, ,
\end{equation}
where $N_c$ is the number of $c_i$. From the assumption that the chiral central charge $c_-$ is a multiple of eight, we find that the topological spin of the $c_i$ indeed satisfies $\theta_c = \theta_\phi^*$. Because $N_{L_f}=\sqrt{N_b}$, it also follows that $N_c=1$, i.e. the anyon $c$ satisfying $\theta_c=\theta_\phi^*$ and $M_{l,c}=\theta_l$ for all $l\in\mathcal{L}_f$ is unique.

The `only if' direction is almost trivial to show. If we assume that the GAfTO = $\mathcal{A}_b\times\{1,\phi,f,\phi f\}$ has a bosonic Lagrangian subgroup, we know that there exists a group of anyons $\mathcal{M}\subset\mathcal{A}_b$ that braid trivially with each other, and have the property that every $a\in\mathcal{A}_b$ which is not in $\mathcal{M}$ braids non-trivially with at least one anyon in $\mathcal{M}$. Because the anyons in $\mathcal{M}$ have trivial mutual braiding, it follows from the ribbon identity that their topological spins form a representation of $\mathcal{M}$: $\theta_{m_1}\theta_{m_2} = \theta_{m_1m_2}$ for all $m_1,m_2\in\mathcal{M}$. From the relation $\theta_{m}^2=M_{m,m}=1$ between topological spin and self-braiding, we also know that $\theta_{m}=\pm 1$ for all $m\in\mathcal{M}$. But this implies that $\mathcal{M}=\mathcal{L}_f$ is a fermionic Lagrangian subgroup of $\mathcal{A}_f=\mathcal{A}_b\times\{1,f\}$.

\subsubsection{Example: gauging U$(1)_4\times\overline{IQH}$}
To illustrate the general result we revisit the example discussed above, i.e. the $\mathcal{A}_f=\mathbb{Z}_4\times\{1,f\}=\{a,a^2,a^3,1\}\times\{1,f\}$ AfTO. Regardless of how we choose the charge vector $t=(2t',-1)$, this AfTO always has a fermionic Lagrangian subgroup given by $\mathcal{L}_f=\{1,a^2\}$. Recall that $\theta_{a^2}=-1$, so this would not be a valid Lagrangian subgroup if the system were bosonic. We first consider the case where the charge vector is given by $t=\left(2,-1\right)^T$. This implies that $\sigma_{xy}=0$, and $q_a=1/2$. This model is known to have a gapped edge, as it is equivalent to a fermionic $\mathbb{Z}_2$ gauge theory \cite{FillingC} (i.e. a fermionic toric code \cite{GuWang}). Let us again denote the fermion parity flux obtained by adiabatic flux insertion as $\phi_{A}$. Because $\sigma_{xy}=0$, $\phi_{A}$ is a boson. Because $a^2$ has charge one (recall that $q_{a^p} = pt'/2$), it also follows that $M_{\phi_{A},a^2}=-1$. From this we see that $\mathcal{L}_b=\{1,a^2f,\phi_{A},\phi_{A}a^2f\}=\{1,a^2f\}\times\{1,\phi_A\}=\mathbb{Z}_2\times\mathbb{Z}_2$ is a bosonic Lagrangian subgroup of the GAfTO. 

Let us now repeat this analysis for the case where $t=\left(0,-1\right)$. This choice of charge vector implies that $\sigma_{xy}=-1$ and $q_a=0$. This fTO is just the stacking of a $\sigma_{xy}=-1$ IQH state, and a purely bosonic U$(1)_4$ topological order (because all anyons in U$(1)_4$ have trivial fermion parity charge). The parity flux obtained by adiabatic flux insertion now has topological spin $\theta_{\phi_{A}}=e^{i\pi\sigma_{xy}/4}=e^{-i\pi/4}$. The flux $\phi_A$ also braids trivially with all anyons in $\mathcal{A}_b=\{1,a,a^2,a^3\}$ because they have zero charge. It is now easy to check that $\mathcal{L}_b=\{1,a\phi_{A},a^2f,a^3\phi_{A}f\}=\mathbb{Z}_4$ is the Lagrangian subgroup of the GAfTO. So in this example, the anyon $a$ plays the role of the special anyon $c$ which occured in the general proof above.

\subsection{Non-Abelian parity flux}

If the fermion parity flux, and therefore also the GfTO, is non-Abelian, we have to use a generalization of Eq. (\ref{kit}) to determine the chiral central charge modulo eight from the bulk data. Writing the quantum dimensions of the anyons in the GfTO order as $d_a$, the general expression for $c_-$ becomes \cite{Kitaev}:

\begin{equation}\label{nonAb}
e^{2\pi i c_-/8} = \frac{1}{\mathcal{D}} \sum_a d_a^2 \theta_a\, ,
\end{equation}
where the total quantum dimension is given by $\mathcal{D}=\sqrt{\sum_a d_a^2}$. In a GAfTO $\mathcal{A}_b\times\{1,\phi,f\}$, the only non-Abelian anyons are $\mathcal{A}_b\times \phi$, and these are therefore the only anyons which have a quantum dimension different from one. From the fusion rule $\phi\times\phi = 1+f$, it follows that $d_\phi = \sqrt{2}$. Applying Eq. (\ref{nonAb}) to a non-Abelian GAfTO, we find

\begin{equation}\label{eqdouble}
e^{2\pi i c_-/8} = \frac{\theta_\phi}{\sqrt{N_b}}\sum_{a\in\mathcal{A}_b} \theta_a
\end{equation}
This is exactly the same expression as the one we obtained for Abelian fermion parity fluxes. 

We can now easily argue that an AfTO with non-Abelian fermion parity fluxes can never have a gapped edge. First, we note that $N_b^{-1/2}\sum_{a\in\mathcal{A}_b}\theta_a$ is always an eighth root of unity. This is because Abelian topological orders have a multi-component U$(1)$ Chern-Simons description, such that the corresponding edge theories are chiral Luttinger liquids with integer chiral central charge \cite{XGWen}. On the other hand, if $\phi$ is non-Abelian, it has a topological spin $\theta_\phi = e^{2\pi i (2n+1)/16}$, where $n$ is an integer \cite{Kitaev}. So from Eq. (\ref{eqdouble}), it follows that $c_-$ is a half odd integer, which means that the edge is always chiral and cannot be gapped.

\section{Conclusions}
In this work we have explored the special structure of gauged Abelian fermionic topological orders, and we have exploited this structure to study both the numerical detection of such phases and the fate of the edge physics under the gauging process. We have outlined a minimal scheme to uniquely identify the AfTO realized by a microscopic lattice Hamiltonian, which does not make use of fermion number conservation symmetry. We have also shown that a gauged fTO can have a gapped bosonic edge to the vacuum if and only if the original, ungauged fTO admits a fermionic gapped edge.

An obvious question is of course how to generalize these results to systems with non-Abelian anyons. The mathematical framework required  to address non-Abelian fermionic topological orders is developed and discussed in Refs. \cite{GuWangWen,TianLan1,TianLan2,Wang,Aasen}, and it is substantially more involved than the simple arguments used in this work. However, the understanding of Abelian systems provides a clear, intuitive picture of the physics involved, and hopefully this will be helpful for a rigorous study of non-Abelian systems. We leave such a study for future work.

\subsubsection{Acknowledgements} I am grateful to Meng Cheng for pointing out theorem 3.13 from Ref. \cite{Drinfeld} to me, and for a collaboration on a previous project which was very useful for the present paper. I also want to thank Mike Zaletel for inspiring discussions which formed the motivation for the present work, and Johannes Motruk for discussions and for pointing me to some important references. During the completion of this work I was supported by the DOE, office of Basic Energy Sciences under contract no. DE-AC02-05-CH11231.

\appendix

\section{Example of momentum polarization in fermionic systems: Chern insulator and topological $p+ip$ superconductor}

In this appendix, we present an example of the application of momentum polarization to fermion systems. We consider a system with $\mathcal{A}_b=1$, described by the translationally invariant, spinless free fermion Hamiltonian on the square lattice

\begin{equation}\label{freefermion}
H = \sum_{\textbf{k}}\psi^\dagger_{\textbf{k}} h(\textbf{k})\psi_{\textbf{k}} \, ,
\end{equation}
where the single particle Hamiltonian is given by

\begin{equation}
h(\textbf{k}) = \textbf{d}(\textbf{k})\cdot \mathbf{\sigma}=\left( \begin{array}{cc} \cos(k_x)+\cos(k_y)-1 & \sin(k_x)+i\sin(k_y) \\  \sin(k_x)-i\sin(k_y) & -\cos(k_x)-\cos(k_y)+1\end{array}\right)
\end{equation}
Importantly, $|\textbf{d}(\textbf{k})|$ is non-zero in the entire Brillouin zone, and the map $\textbf{k}\rightarrow \textbf{d}(\textbf{k})/|\textbf{d}(\textbf{k})|$ covers the unit sphere once. This implies that the gapped bands of $h(\textbf{k})$ have Chern number $\pm 1$. 

\begin{figure}
\begin{center}
a)
\includegraphics[scale=0.27]{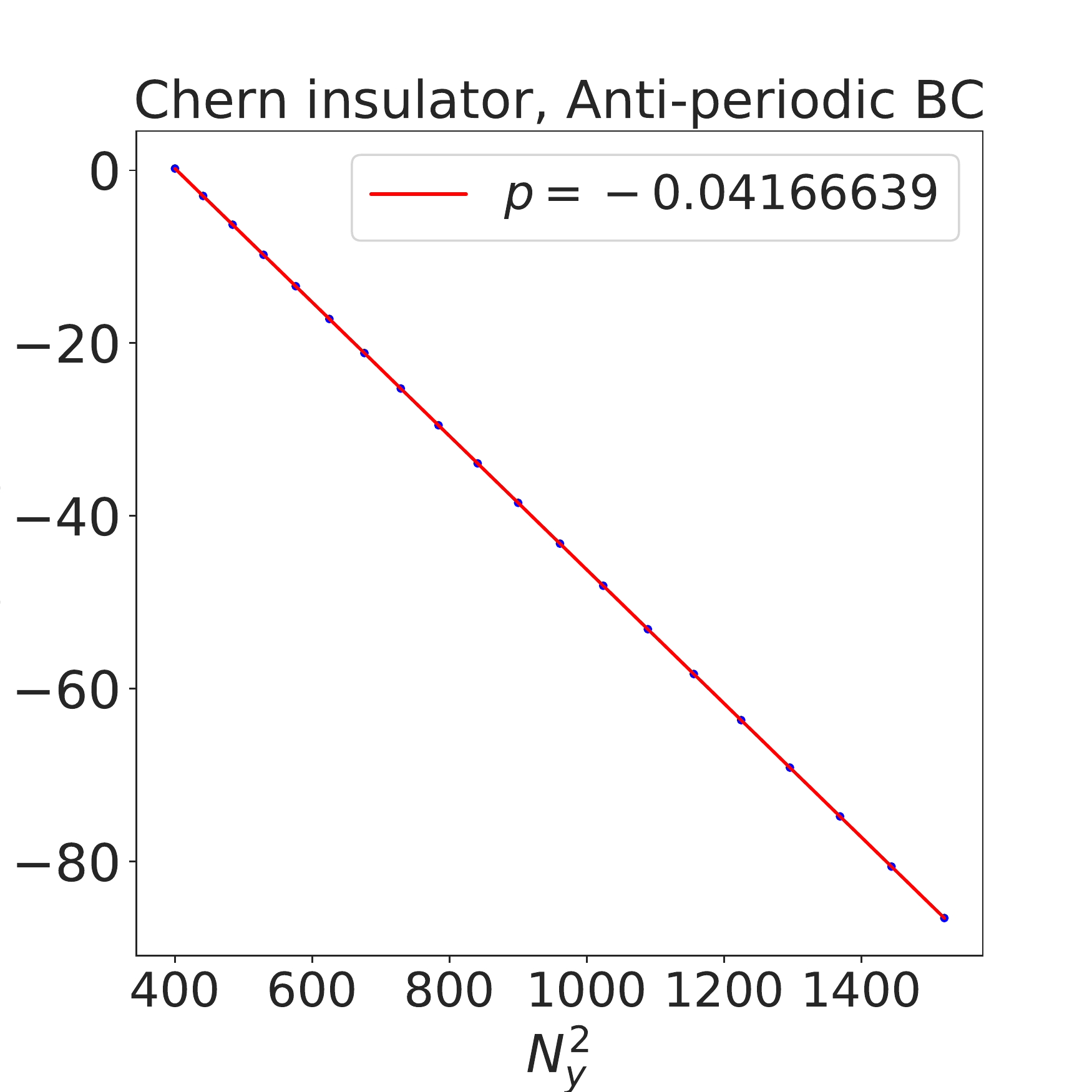}
b)
\includegraphics[scale=0.27]{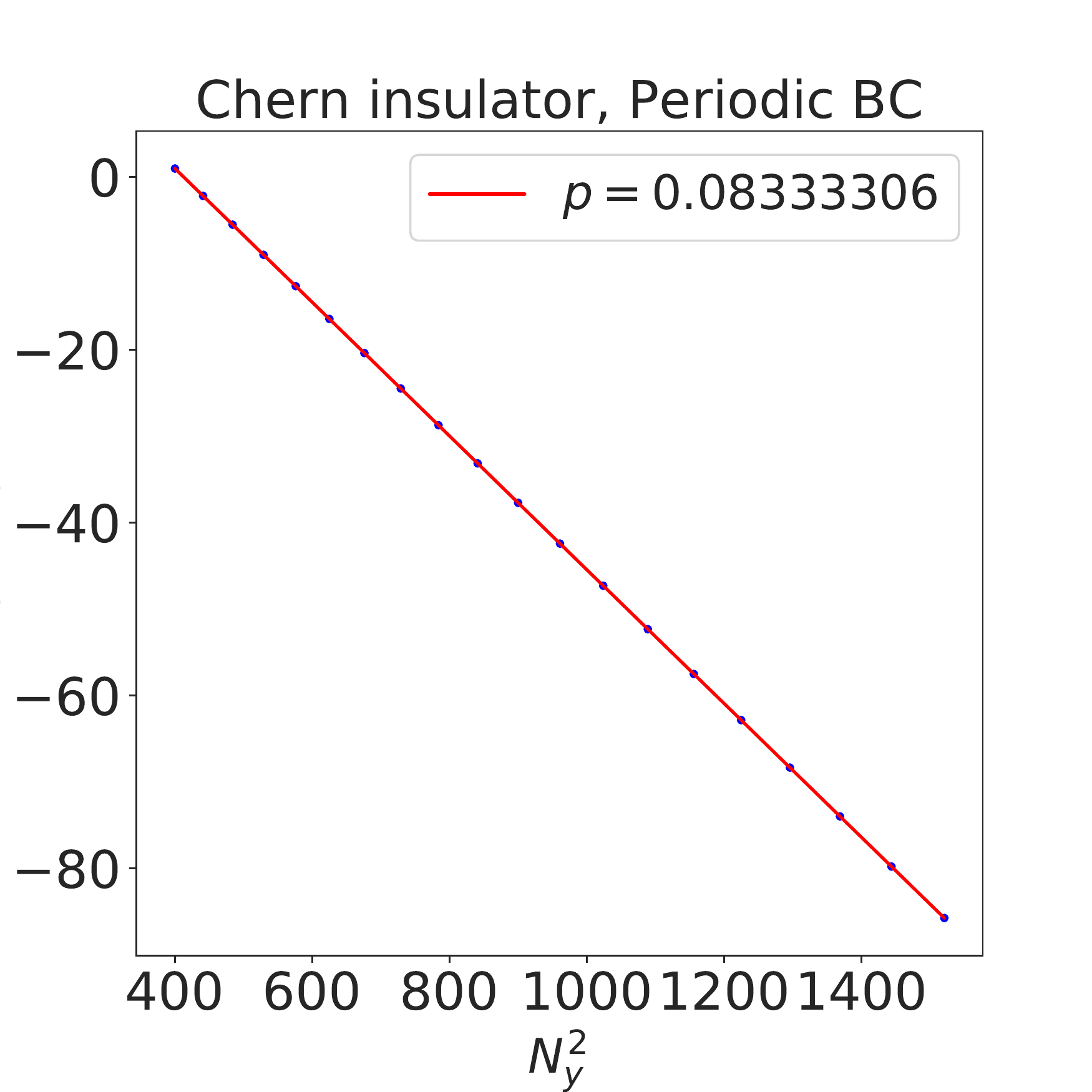}
c)
\includegraphics[scale=0.27]{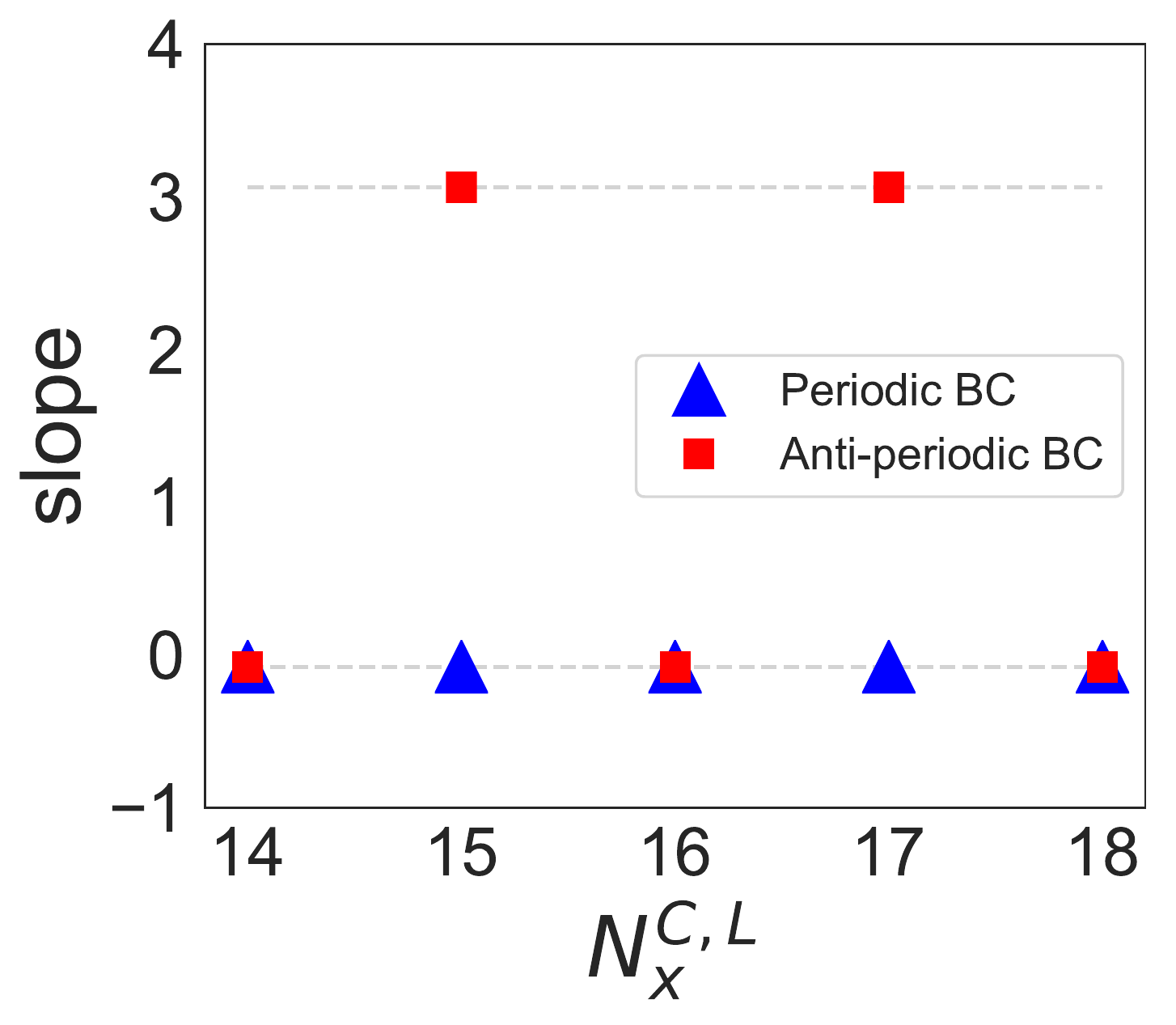}
\caption{(a) The momentum polarization for a Chern insulator on a cylinder of size $N_x\times N_y$ with fixed length $N_x =30$. $N_y \theta_C(N_y)$ is plotted as a function of $N_y^2$, where $\theta_C(N_y)$ is defined as $\langle \psi_f[C,1]|\tilde{T}^{C,L}_y|\psi_f[C,1]\rangle = |\langle \psi_f[C,1]|\tilde{T}^{C,L}_y|\psi_f[C,1]\rangle| \exp(i\theta_C(N_y))$. $N_y\theta_C(N_y)$ was determined numerically as $N_y\theta_C(N_y) = \ln( \exp(iN_y\theta_C(N_y))$ (see the discussion in the main text for why this is important). The intercept $2\pi p$ of the linear fit to the data points gives access to the chiral central charge via the relation $p=-c_-/24 = -1/24 \approx -0.041666$ mod 1. The entanglement cut was chosen such that $N_x^{C,L}=16$. (b) Same as in (a), but now in the periodic sector. The intercept $2\pi p$ is determined by $p = h_\phi -c_-/24 = 1/8 - 1/24 = 1/12 \approx 0.08333$ mod 1. (c) The slope of the linear fits in $(a)$ and $(b)$ as a function of $N_x^{C,L}$.}\label{intercept}
\end{center}
\end{figure}

The free fermion Hamiltonian (\ref{freefermion}) has two possible interpretations. Either we interpret it as a charge conserving model with two orbitals $A$ and $B$ on each site, in which case the vector of annihilation operators is given by

\begin{eqnarray}
\psi_{\textbf{k}} & = &  \left(\begin{array}{c} c_{A,\textbf{k}} \\ c_{B,\textbf{k}} \end{array}\right)\, ,
\end{eqnarray}
or we interpret $H$ as a superconductor, in which case $\psi_{\textbf{k}}$ is a Nambu spinor given by

\begin{eqnarray}
\psi_{\textbf{k}} & = &  \left(\begin{array}{c} c_{\textbf{k}} \\ c^\dagger_{-\textbf{k}} \end{array}\right)
\end{eqnarray}
In the first interpretation, the non-zero Chern number of the bands implies that the system is an anomalous Hall (or Chern) insulator. In the superconducting case, the system is a topological or weak-pairing $p+ip$ superconductor \cite{ReadGreen}.

For free fermion systems there is a straightforward way to obtain the eigenvalues of the reduced density matrix corresponding to some spatial region \cite{Peschel}. First, we define the one-particle reduced density matrix as

\begin{equation}
C_{(j,k_y),(j',k_y')} = \langle c^\dagger_{(j,k_y)} c_{(j',k_y')}\rangle = \delta_{k_y,k_y'}\langle c^\dagger_{(j,k_y)} c_{(j',k_y)}\rangle\, ,
\end{equation}
where $j$ is a spatial index along the axis of the cylinder, and $k_y$ is the momentum along the periodic direction. Next, we take the subblock of $C_{(j,k_y),(j',k_y)}$ where both $j$ and $j'$ lie on the left of the entanglement cut, and calculate the eigenvalues $\zeta_{k_y,n}$ of that subblock. The eigenvalues of the reduced density matrix corresponding to the left half of the cylinder are then labeled by a set of occupation numbers $n_{k_y,n}\in \{0,1\}$, and are given by \cite{Peschel}

\begin{equation}
\lambda[\{n_{k_y,n}\}] = \prod_{k_y,n} \left(\zeta_{k_y,n}\right)^{n_{k_y,n}}(1-\zeta_{k_y,n})^{1-n_{k_y,n}}
\end{equation}
Using this expression for the eigenvalues of the reduced density matrix, the momentum polarizations for the Chern insulator in the anti-periodic and periodic sectors can readily be obtained as \cite{TuZhang,Alexandradinata}

\begin{eqnarray}
\langle \psi_f[C,1]|\tilde{T}^{C,L}_y|\psi_f[C,1]\rangle & = & \prod_{-\pi< k_y \leq\pi }\prod_n\left((e^{ik_y}-1)\zeta_{k_y,n}+1 \right)\\
\langle \psi_f[C,\phi]|T^{C,L}_y|\psi_f[C,\phi]\rangle  & = & \prod_{-\pi< k_y \leq\pi }\prod_n\left((e^{ik_y}-1)\zeta_{k_y,n}+1 \right)\, ,
\end{eqnarray}
where in the anti-periodic (periodic) sector $k_y = \frac{2\pi}{N_y}\left(j+\frac{1}{2}\right)-\pi, \, j\in \{0,\dots,N_y-1\}$ for $N_y$ even (odd), and $k_y = \frac{2\pi}{N_y}j - \pi, \, j\in \{0,\dots,N_y-1\}$ for $N_y$ odd (even). For the $p+ip$ superconductor, the expressions for the momentum polarizations are very similar:

\begin{eqnarray}
\langle \psi_f[C,1]|\tilde{T}^{C,L}_y|\psi_f[C,1]\rangle & = & \prod_{0\leq k_y\leq \pi }\prod_n \left((e^{ik_y}-1)\zeta_{k_y,n}+1 \right) \\ 
\langle \psi_f[C,\phi]|T^{C,L}_y|\psi_f[C,\phi]\rangle & = &  \prod_{0\leq k_y\leq \pi }\prod_n \left((e^{ik_y}-1)\zeta_{k_y,n}+1 \right)\, ,
\end{eqnarray}
and the only difference is in the range of $k_y$. Using the particle-hole symmetry $\sigma^xh(-\textbf{k})^*\sigma^x = -h(\textbf{k})$, which implies that

\begin{equation}\label{PH}
\zeta_{k_y,n} = 1-\zeta_{-k_y,n}\, ,
\end{equation}	
one can find an exact relation between the momentum polarizations of the Chern insulator and the $p+ip$ superconductor. From Eq. (\ref{PH}), one finds that

\begin{equation}
\prod_{0<k_y<\pi ,n}\left( (e^{-ik_y} -1)\zeta_{-k_y,n} + 1\right) = \prod_{0<k_y<\pi,n} e^{-ik_y}\left( (e^{ik_y} -1)\zeta_{k_y,n} + 1\right) 
\end{equation}
This implies that the momentum polarizations of the Chern insulator are the square of those of the $p+ip$ superconductor, up to a factor $\prod_{0<k_y<\pi,n} e^{-ik_y}$. This factor only changes the term in the exponent of the momentum polarization which is linear in $N_y$, implying that both the chiral central charge and the topological spin of the fermion parity flux differ by a factor of two between the Chern insulator and the $p+ip$ superconductor. This is of course consistent with the known values $c_- = 1$ and $h_\phi = 1/8$ for the Chern insulator and $c_-=1/2$ and $h_\phi = 1/16$ for the $p+ip$ superconductor.

Let us write the momentum polarization in the anti-periodic sector as $\langle \psi_f[C,1]|\tilde{T}^{C,L}_y|\psi_f[C,1]\rangle = |\langle \psi_f[C,1]|\tilde{T}^{C,L}_y|\psi_f[C,1]\rangle| \exp(i\theta_C(N_y))$. In Fig. \ref{intercept} (a), $N_y\theta(N_y)=\ln(\exp(iN_y\theta_C(N_y))$ is plotted as a function of $N_y^2$ for the Chern insulator. According to the discussion above, $\theta_C(N_y)$ is given by

\begin{equation}
\theta_C(N_y) = -\frac{2\pi i}{N_y}  \frac{c_-}{24} - \alpha_C N_y\, ,
\end{equation} 
where $\Delta\alpha_C = \pi$ because the Chern insulator has an odd fermion parity per site ($\sigma = 1$). This means that $N_y\theta_C(N_y)$ has a linear dependence on $N_y^2$, and the intercept is determined by the chiral central charge. Because $\Delta\alpha_C = \pi$, the slopes for two neighboring cuts differ by $\pi$. This dependence of the slope on the choice of cut is also found numerically, as shown in Fig. \ref{intercept} (c). 

In the periodic sector, $\theta_C(N_y)$ is defined as $\langle \psi_f[C,\phi]|T^{C,L}_y|\psi_f[C,\phi]\rangle = |\langle \psi_f[C,\phi]|T^{C,L}_y|\psi_f[C,\phi]\rangle| \exp(i\theta_C(N_y))$, and it takes the form

\begin{equation}\label{thetac}
\theta_C(N_y) = \frac{2\pi i}{N_y} \left(h_{\phi } -\frac{c_-}{24}\right) -\pi N_x^{C,L}- \alpha_C N_y \, ,
\end{equation}
where again $\Delta\alpha_C=\pi$. From this equation it is clear that $e^{i\theta_C(N_y)}$ differs by a factor of $(-1)^{N_y+1}$ for two neighboring cuts. So if we compute $N_y\theta_C(N_y)$ numerically by taking the logarithm of $e^{iN_y\theta_C(N_y)}$, then the dependence on the cut of the terms in Eq. (\ref{thetac}) which are constant and linear in the circumference $N_y$ will not show up. This means that $N_y\theta_C(N_y)$ computed in his way has a linear dependence on $N_y^2$, with a slope which does not depend on the choice of cut. In Fig. \ref{intercept} (b), this linear dependence of $N_y\theta_C(N_y)$ on $N_y^2$ is plotted. From the intercept of this line we can find the topological spin of the parity flux. In Fig. \ref{intercept} (c), the independence of the slope on the choice of cut is shown. However, if we would instead calculate $N_y\theta_C(N_y)$ as $N_y\theta_C(N_y)= N_y \ln(\exp(i\theta_C(N_y))$, then it follows from Eq. (\ref{thetac}) that $N_y\theta_C(N_y)$ would not be linear in $N_y^2$ for odd $N_x^{C,L}$. We have verified that this is indeed the case, and that for odd $N_x^{C,L}$, $N_y\theta_C(N_y) = N_y \ln (\exp(i\theta_C(N_y)+\pi(N_y+1)))$ is linear in $N_y^2$, with the same slope as for even $N_x^{C,L}$. This provides a non-trivial consistency check on the generalized expressions for the momentum polarizations in Eqs. (\ref{mompolAP}) and (\ref{mompolP}).

\section*{References}

\bibliographystyle{iopart-num}
\bibliography{bibliography}

\end{document}